\documentclass[prl,twocolumn,superscriptaddress,showpacs,nobalancelastpage,amsmath,amssymb]{revtex4}
\usepackage{graphicx}% Include figure files
\usepackage{bm}% bold math

\begin{document}
\date{\today}
%opening
\title{
Quantum anomalous Hall effect in Mn doped HgTe quantum wells
}

\author{A.\ Budewitz,$^1$ K.\ Bendias,$^1$ P.\ Leubner,$^1$ T. Khouri,$^2$ \\ S.\ Shamim,$^1$ S. Wiedmann,$^2$ H.\ Buhmann,$^1$ and L.W.\ Molenkamp}

\affiliation{Faculty of Physics and Astrophysics, University of W\"urzburg, Am Hubland, 97074 W\"urzburg, Germany
\\
$^2$High Field Magnet Laboratory (HFML-EMFL), Radboud University, Nijmegen, the Netherlands}

\begin{abstract} 
\textbf{
Magnetotransport measurements are presented on paramagnetic (Hg,Mn)Te quantum wells (QWs) with an inverted band structure. Gate-voltage controlled density dependent measurements reveal an unusual behavior in the transition regime from n- to p-type conductance: A very small magnetic field of approximately 70 mT is sufficient to induce a transition into the $\nu = -1$ quantum Hall state, which extends up to at least 10 Tesla. The onset field value remains constant for a unexpectedly wide gate-voltage range. Based on temperature and angle-dependent magnetic field measurements we show that the unusual behavior results from the realization of the quantum anomalous Hall state in these magnetically doped QWs.
}
\end{abstract}

\maketitle

%\section{Introduction}
%{\em Introduction} -- 
The experimental discovery of the quantum spin Hall (QSH) effect \cite{Koenig} in HgTe quantum wells (QWs) with an inverted band structure raised the interest in the electronic properties of two-dimensional (2D) topological insulators (TI). At zero magnetic field, transport occurs non-local and spin-polarized \cite{Roth,Bruene}. However, it turns out that quantized longitudinal conductance is only observable in micrometer scale devices. Samples exceeding these dimensions exhibit backscattering (cf.\  suppl.\ mat.\ Ref.\ \cite{Roth}). 

\begin{figure}[t]
\includegraphics[width=75mm]{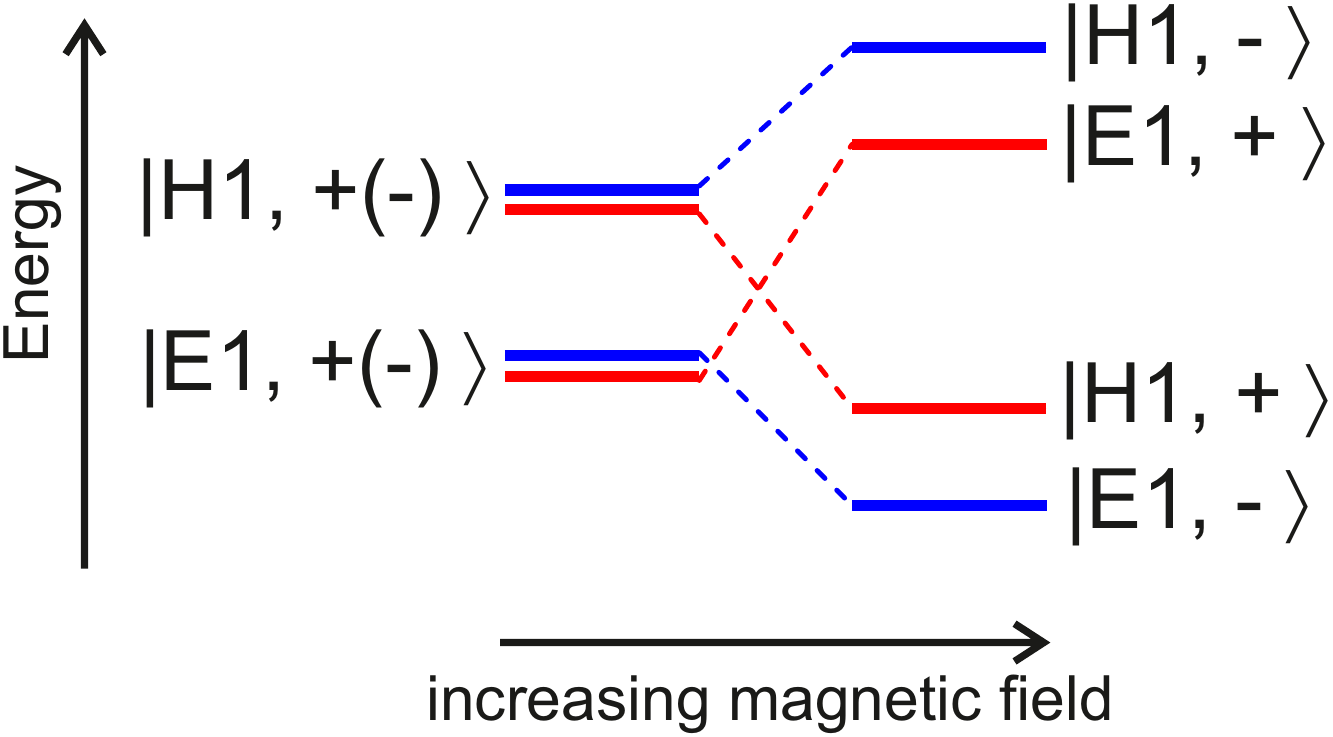}
\caption{Evolution of the band structure in Mn-doped inverted QW with increasing magnetic field.}
\label{Figure_1}
\end{figure}

Shortly after the QSH discovery, C.X. Liu {\it et al.} \cite{Liu} suggested that the QSH state can be transferred into a QH state with transverse quantized conductance. In principle, this so-called quantized anomalous Hall (QAH) state can be realized at zero magnetic field in ferromagnetic 2D TIs, in which the exchange field lifts the spin degeneracy of the bulk state. This causes the bulk band gap to increase for one spin orientation while it decreases for the other, eventually reversing into a normal order for sufficient depending on the field strength. The resulting energy band evolution is schematically shown in Fig.\ \ref{Figure_1}. 
%%%  Figure 1   %%%
%
%\vfill
%
In this limit, the sample exhibits only one chiral edge channel, which gives rise for the QAH effect. It turns out that the exchange field in Mn-doped HgTe QW structures is not strong enough to drive the material to become ferromagnetic as proposed by Liu \textit{et al.} \cite{Liu}. 
The situation is quite different for magnetically doped Bi-compounds, where the QAH state could be convincingly demonstrated \cite{Chang1, Chang2, Grauer}. In the present communication we show that paramagnetic (Hg,Mn)Te QW structures also do support the QAH effect, but need a small magnetic field to ensure sufficient spin alignment.

We present magneto-transport studies of Mn-doped HgTe QWs at different temperatures and magnetic fields. We observe a transition from the QSH  state into the $\nu = -1$ chiral anomalous QH state at very low magnetic fields ($B < 70$~mT), which is related to the paramagnetic ordering of the Mn-ions. This interpretation is supported by temperature dependent measurements and additionally, by exploring the influence of an in-plane magnetic field (which only affects the Zeeman but not the orbital Hall contribution) on magnetically and non-magnetically doped QWs. 

%{\em Experiment} --

We studied a series of Hg$_{\mathrm{1-x}}$Mn$_\mathrm{x}$Te quantum well structures, grown by molecular beam epitaxy. The Mn concentration varies between 0.5 to $4 \%$ for QWs of thicknesses between $d = 6$ nm to $d = 11$ nm covering the regime from normal to inverted band structure ordering. The critical QW thickness $d_\mathrm{c}$ depends on the Mn concentration $(x)$ and ranges from $d_\mathrm{c}(x = 0) = 6.3$ nm to $d_\mathrm{c}(x = 0.04) \approx 12$ nm (cf.\ \cite{supplmat} Fig.\ S1). Since Mn substitutes Hg isoelectrically, Mn atoms introduce neither additional doping, nor do they act as defects. The QWs are embedded between 25 to 50 nm thick Hg$_{\mathrm{0.3}}$Cd$_{\mathrm{0.7}}$Te barrier layers. Standard Hall bars were fabricated by dry etching. Au gate electrodes were evaporated onto a 110 nm thick $\mathrm{SiO}_2$/SiN multilayer insulator stack, which allows for an effective control of the charge carrier density in the QW. 
%
%%%  Figure 2   %%%
%

\begin{figure*}[!ht]
\includegraphics[width=150mm]{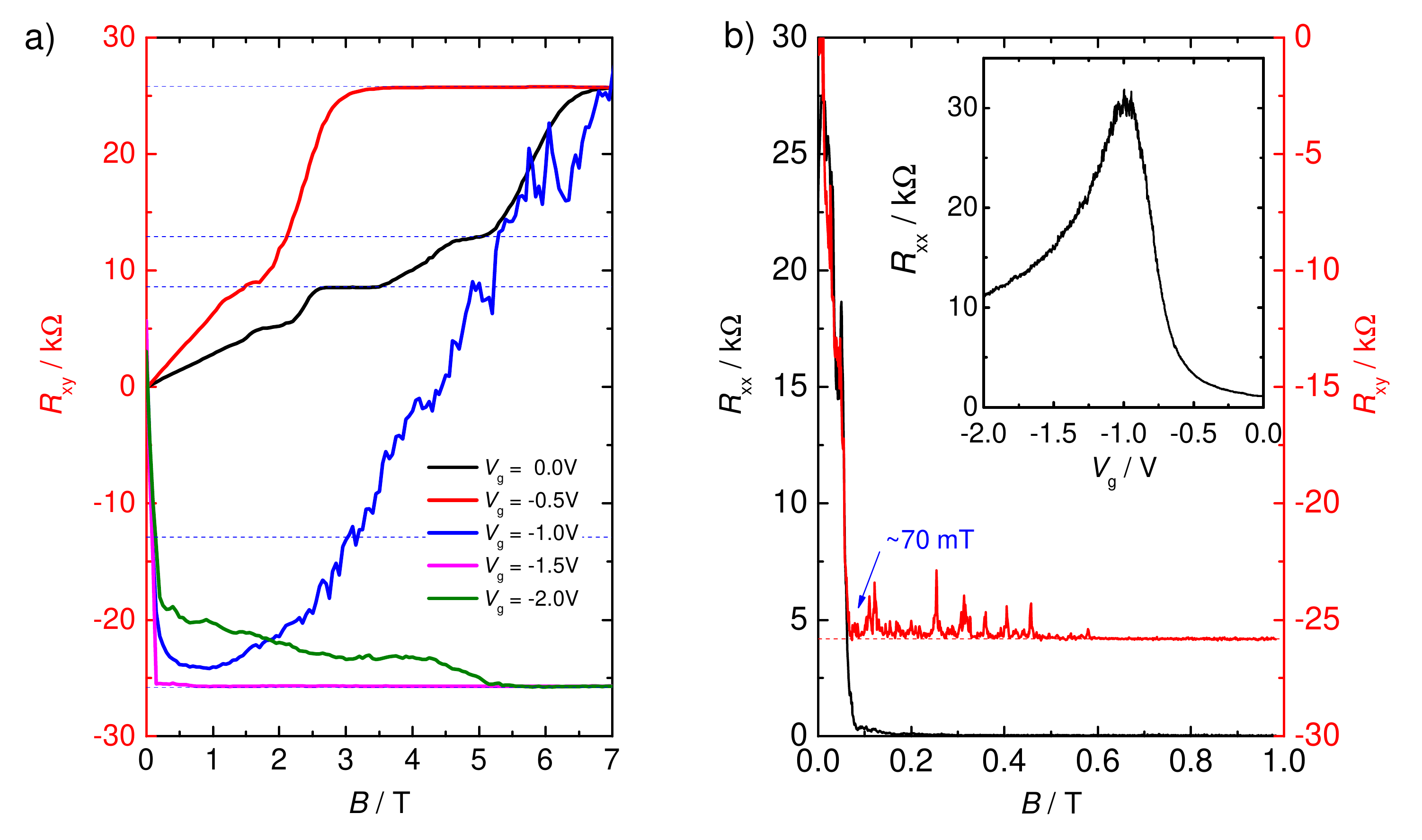}
\caption{Magnetic field and gate voltage dependent measurement for a $10$~nm magnetically doped (Hg,Mn)Te quantum well with $2\%$ Mn at a base temperature of $10$ mK. (a) Hall resistance $R_{\mathrm{xy}}$ for different gate voltages, (b) $R_{\mathrm{xx}}$ and $R_{\mathrm{xy}}$ for $V_\mathrm{g} = -1.2$~V. The inset of (b) shows the gate voltage dependent longitudinal resistance $R_{\mathrm{xx}}$ at $B = 0~\mathrm{T}$.}
\label{Figure_2}
\end{figure*}
Magneto-transport measurements were carried out in different cryostats using ac-measurement techniques for temperatures down to 0.01 K. 

In this work, we focus on results from a 10 nm wide (Hg,Mn)Te QW with a Mn concentration of about $2\%$ ($d_\mathrm{c} \approx 9$ nm) and present data for a $30 \times 10\ \mathrm{\mu m}^2$ device. For this QW width and Mn concentration the key feature, the $\nu = -1$ quantization, shows the earliest onset in magnetic field. Data for the other samples are provided as supplementary material \cite{supplmat}. 

Fig.\ 2a shows the measured Hall resistance ($R_{\mathrm{xy}}$) for different gate voltages ($V_\mathrm{g}$). (The corresponding longitudinal resitances $R_{\mathrm{xx}}$ are shown in Fig.~S2 of \cite{supplmat}.) It indicates a transition from the n- to the p-conducting regime for $V_\mathrm{g} \approx -1$~V. Correspondingly, the gate voltage dependent longitudinal resistance, $R_{\mathrm{xx}}$, at $B = 0$ exhibits a maximum around $V_\mathrm{g}= -1$~V (inset Fig.\ 2b), which is associated with the bulk band gap. In this regime, current is carried by helical edge channels forming the QSH state. 
The maximum in $R_{\mathrm{xx}}$ does not show the expected quantized value which we ascribe to the dimensions of the device. (We will address the issue  of conductance quantization in (Hg,Mn)Te in a separate publication.)

Surprisingly, at very small magnetic fields, when the Fermi level approaches the bulk gap, the conductance of the Hall bar immediately changes from n- to p-type and a $\nu = -1$ plateau forms almost instantly. In Fig.~S1 \cite{supplmat} of the supplementary section we show data on the differential conductance, presenting the color-coded Landau level fan charts, for all measured samples. It is obvious that the peculiar behavior of Fig.\ 2a is unique for this QW width and Mn-concentration. Especially for non-inverted and non-magnetic samples a clear insulating regime characterizes the transition regime from n- to p-conductance, which is absent in Fig.\ 2a. 

Fig.\ 2b shows the Hall resistance for $V_\mathrm{g} = -1.2$~V. For this gate voltage the $\nu = -1$ plateau is well developed, with an onset at magnetic fields as low as $B \approx 70$~mT. It is remarkable that also the $\nu= -1$ plateau extends to very high magnetic fields exceeding the presented regime of $B = 0$ to 7~T (Fig.\ 2a, $V_\mathrm{g} = 1.5$~V) while the onset still remains at around 70~mT. The latter is in clear contrast to non-magnetic QWs. The earliest onset for non-magnetic QW is between 0.5 and 1~T (cf.\ \cite{supplmat} Fig.~S1(I), and Ref.\ \cite{KoenigJPSJ}). However, this onset shifts rapidly towards higher magnetic field with decreasing gate voltage. For the present (Hg,Mn)Te sample deviations from the early $\nu = -1$ quantization are only observed for gate voltages more negative than $-1.5$~V (cf.\ Fig.\ 2a). At $-2.0$~V the onset is already shifted to $B \approx 5$~T, now resembling the usual QH effect of a HgTe QW in the p-conducting regime (cf.\ \cite{supplmat} Fig.\ S1).

%Discussion 1

In order to explain the experimental observation we refer to the model presented by Liu \textit{et al.}  \cite{Liu} (cf.\ Fig.\ 1). At low temperatures and for Mn-concentrations below $15\%$, (Hg,Mn)Te is paramagnetic \cite{Brandt}. The g-factor for the electron- and hole-like states have opposite signs and consequently the same holds for the magnetic field induced spin-splitting, $G_{i} = g_{i}\langle S \rangle$ with $i = E,H$ and $\langle S \rangle$ the spin polarization of Mn. In the presence of a magnetic field, the $|E,+\rangle$ ($|E,-\rangle$) and the $|H,+\rangle$ ($|H,-\rangle$) states experience an opposite energy shift. Since in the present case $G_E < 0$ and $G_H > 0$, already at small magnetic fields we reach a situation where the inversion between the $|E,+\rangle$ and $|H,+\rangle$ states is undone, while the $|E,-\rangle$ and $|H,-\rangle$ states remain inverted and form a chiral QAH edge state (cf.\ Fig.\ 1). For the sample of Fig.\ 2a this transition occurs at magnetic fields as low as $70$ mT. We surmise that for this sample (QW width $d = 10$~nm and Mn-concentration, $x = 2\%$) the magnetic field induced spin splitting and the bulk band gap at zero magnetic field  ($E_{\rm gap} = 12$~meV, inferred from band structure calculations \cite{Novik}) are nearly perfectly matched for the observation of a low field $\nu = -1$ QAH state.

The influence of magnetic doping on the g-factor is already visible in the sequence of QH plateaus in the n-conducting regime for this sample (Fig.\ 2a). While for $V_g = 0$~V a complete even-odd sequence of QH plateaus is observed the even plateaus are missing at slightly lower carrier density ($V_g = -0.5$~V).
This behavior is in agreement with model eight band \textbf{k}$\cdot$\textbf{p} calculations and experimentally observed beating pattern in the corresponding Shubnikov-de Haas oscillations \cite{Novik,Gui}. Similar g-factor induced Landau level crossings are observed for all other magnetic QWs \cite{supplmat}.
%
%%%  Figure 3   %%%
\begin{figure}[!h]
\includegraphics[width=75mm]{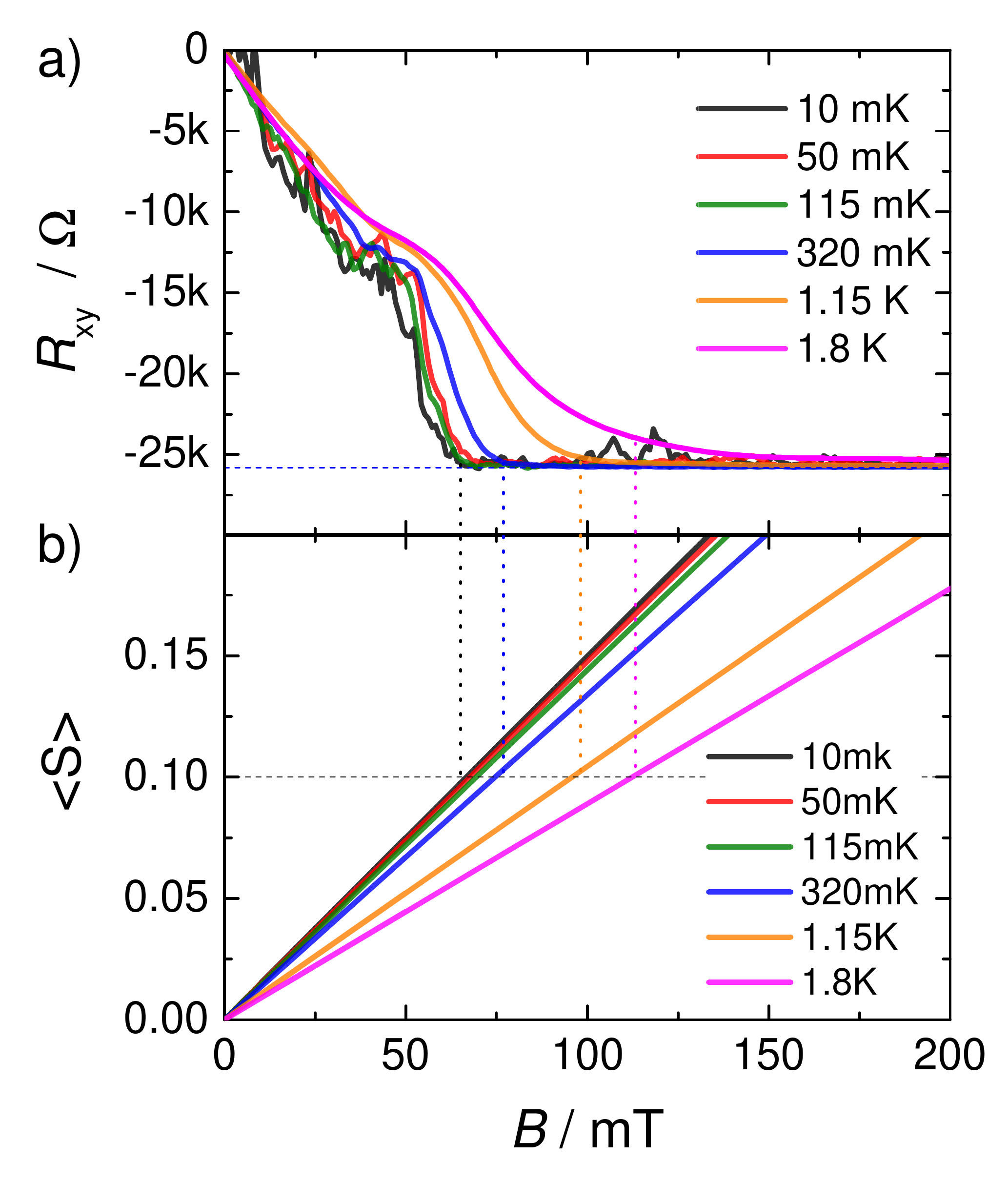}
\caption{ (a) The Hall resistance $R_{\mathrm{xy}}$ as function of perpendicular magnetic field $B$ for the $10$ nm (Hg,Mn)Te quantum well at $V_g = -1.2$~V for different temperatures ranging from $10$ mK to $1.8$ K. The horizontal dashed blue line indicate the quantized Hall resistance $R_{\mathrm{xy}} = -25.8$ k$\Omega$. (b) The spin polarization of the Mn ions as a function of $B$ for different temperatures (calculated using Eq.\ \ref{Magnetization}).}
\label{Figure_3}
\end{figure}

Since the transition into the $\nu = -1$-state is related to the spin polarization of the Mn ions, the onset of the plateau should reflect the magnetization of Mn and thus strongly depend on temperature. Fig.\ \ref{Figure_3} compares the Hall resistance for different temperatures (a) with the corresponding Mn spin polarization $\langle S \rangle$ (b), which has been calculated using
%
%%%   Equation 1   %%%
%
\begin{eqnarray}
{\langle\cal{S}\rangle} = -S_0 B_{5/2} \left( \frac{5 g_{\rm Mn} \mu_B B}{2k_B(T+T_0)} \right),
\label{Magnetization} 
\end{eqnarray}
where ${\cal{S}}_0 =5/2$ denotes the Mn spin, $g_{\mathrm{Mn}}=2$ is the g-factor of Mn, $T_0=2.6\ \mathrm{K}$ is a compensation term to account for cluster formation and the antiferromagnetic interaction between Mn ions and the $B_{5/2}$ represents the Brillouin function for spin $5/2$ \cite{Winkler, Novik}.
We note that the onset of the $\nu = -1$-plateau appears always for the same value of the spin polarization i.e., $\langle S \rangle \approx 0.1$ (indicated by dashed lines in Fig.\ \ref{Figure_3}a and b). This confirms the conjecture that a certain magnetization is needed to lift the bulk band inversion for one kind of spins and to create a sufficient bulk band gap to form a stable chiral QAH edge channel. 
%higher temp. T > 2 K magnetic phase transition: S -> P!!!

To further confirm that the $\nu = -1$ QH state originates from the paramagnetic ordering of the Mn-ions, we have carried out measurements while varying the angle of the magnetic field with respect to the plane of the QW. In contrast to the QH effect, which depends only on the amplitude of the magnetic field perpendicular to the plane of the two-dimensional electron gas, paramagnetic Mn-ordering is sensitive to the total magnetic field. 

Accordingly, magnetic phase diagram for the QAH state of (Hg,Mn)Te, calculated by Hsu \textit{et al.} \cite{Hsu}, shows that the maximum perpendicular field $B_\mathrm{z}$ for the $\nu = 1$ plateau in the QAH state decreases with increasing in-plane magnetic field ($B_\mathrm{x, eff}$) while the ordinary QH state depends only on $B_\mathrm{z, eff}$. We have experimentally investigated the evolution of the $\nu=-1$ plateau as a function of angle between the normal to the plane of the QW and the magnetic field vector for the Mn-doped and non-magnetic QW. 
%
%%%  Figure 4   %%%
%
\begin{figure*}[t!]
\includegraphics[width=150mm]{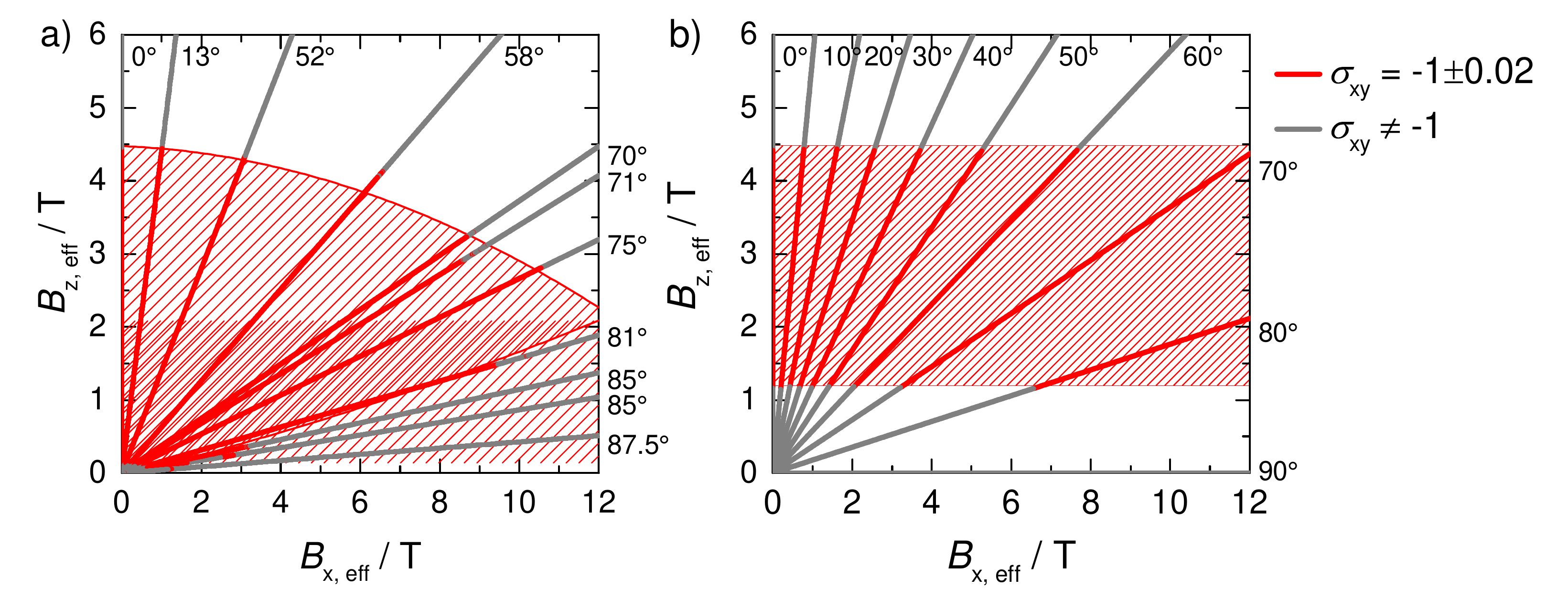}
\caption{Stability range of the $\nu = -1$ plateau $\sigma_\mathrm{xy}$ (red line: $\sigma_{\mathrm{xy}} = -1\pm 0.02$ $e^2/h$ and black: else) for different tilt angles between the normal (0$^\circ$) and the plane (90$^\circ$) of the QW, measured at $V_g = -1.2$~V when the $\nu = 1$ plateau shows the earliest onset. The red area is a guide for the eye which highlights the regime where $\nu -1$ quantization is observed. (a) 10~nm HgTe quantum well (QW) containing $2\%$ Mn at $T = 300$~mK and (b) a $7.5$~nm non-magnetic HgTe QW at $T=1.8$~K.}
\label{Figure_4}
\end{figure*}

Fig.\ 4 shows the stability range of $\nu = -1$ Hall plateau ($\sigma_{\mathrm{xy}}$) for the magnetic QW at a gate voltage of $-1.2$~V and for the non-magnetic QW with comparable carrier density. The red line indicates the magnetic field range where the conductance value is close to $-e^2/h$. One observes that the maximum $B_\mathrm{z}$ for the $\nu=-1$ plateau indeed decreases with increasing in-plane magnetic field for Mn-doped QW (Fig.\ \ref{Figure_4}a), while the QH plateau for the non-magnetic QW is affected only by the strength of $B_\mathrm{z}$ (Fig.\ \ref{Figure_4}b). In further agreement with the phase diagram in Ref. \cite{Hsu}, the onset of the $\nu = -1$ plateau in the Mn-doped case is shifted to higher $B_{\rm z}$ with increasing $B_{\rm x}$ which is caused by the interplay between Zeeman and exchange coupling. The data in Fig.~4 thus constitute further evidence that we indeed observe a quantum anomalous Hall state in our Mn-doped QW. (The experimental raw data can be found in Fig.~S4 of the supplementary section \cite{supplmat}).

In summary, we have carried out detailed magneto-transport measurements on magnetically ordered (Hg,Mn)Te quantum wells and demonstrated that the transition from quantum spin Hall state into the quantum anomalous Hall state can be observed for a specific QW width and Mn-concentration at very low magnetic fields. The temperature and angle dependent magneto-transport measurements confirm the formation of the quantum anomalous Hall state by Mn doping.

\acknowledgments

We thank C. X. Liu for fruitful discussions and acknowledge the financial support from the German Research Foundation (The Leibniz Program, Sonderforschungsbereich 1170 ‘Tocotronics’ and Schwer- punktprogramm 1666), the EU ERC-AG program (Project 3-TOP), the Elitenetzwerk Bayern IDK ‘Topologische Isolatoren’ and the Helmoltz Foundation (VITI). Parts of this work were performed at the High Field Magnet Laboratory Radboud University/Fundamental Research on Matter, member of the European Magnetic Field Laboratory (EMFL) and is part of the research programme of the Foundation for Fundamental Research on Matter (FOM), which is part of the Netherlands Organisation for Scientific Research (NWO). 

\end{document}